\documentclass{article}
\usepackage{spconf,amsmath,graphicx}
\usepackage{amsmath,amsfonts,amssymb,pifont}
\usepackage{adjustbox}
\usepackage{comment}
\usepackage{hyperref}
\usepackage{caption}
\usepackage{array}
\usepackage{xcolor}

\usepackage{xparse}

\usepackage{multirow,makecell}
\usepackage{setspace}
\usepackage{bbding}
\usepackage{relsize}
\usepackage{wrapfig}

\makeatletter
\newcommand*{\@rowstyle}{}
\newcommand*{\rowstyle}[1]{
  \gdef\@rowstyle{#1}%
  \@rowstyle\ignorespaces%
}
\newcolumntype{=}{
  >{\gdef\@rowstyle{}}%
}
\newcolumntype{+}{
  >{\@rowstyle}%
}
\makeatother

\title{Training Audio Captioning Models without Audio \vspace{-7pt}}
\name{\relsize{-0.6} Soham Deshmukh$^{1}$, Benjamin Elizalde$^{1}$, Dimitra Emmanouilidou$^{1}$, Bhiksha Raj$^{2}$, Rita Singh$^{2}$, Huaming Wang$^{1}$ \vspace{-8.5pt} }
\address{$^1$Microsoft, $^2$Carnegie Mellon University \\
\{sdeshmukh, benjaminm, diemmano, huawang\}@microsoft.com, \{bhiksha, rsingh\}@cs.cmu.edu \vspace{-8pt}}

\begin{document}

%
\maketitle
\begin{abstract}
\vspace{-0.1in}
Automated Audio Captioning (AAC) is the task of generating natural language descriptions given an audio stream. A typical AAC system requires manually curated training data of audio segments and corresponding text caption annotations. The creation of these audio-caption pairs is costly, resulting in general data scarcity for the task. In this work, we address this major limitation and propose an approach to train AAC systems using only text. Our approach leverages the multimodal space of contrastively trained audio-text models, such as CLAP. During training, a decoder generates captions conditioned on the pretrained CLAP text encoder. During inference, the text encoder is replaced with the pretrained CLAP audio encoder. To bridge the modality gap between text and audio embeddings, we propose the use of noise injection or a learnable adapter, during training. We find that the proposed text-only framework performs competitively with state-of-the-art models trained with paired audio, showing that efficient text-to-audio transfer is possible. Finally, we showcase both stylized audio captioning and caption enrichment while training without audio or human-created text captions.

\end{abstract}
\vspace{-0.1in}
\section{Introduction}\label{sec:intro}
\vspace{-0.1in}
\begin{figure*}[ht]
   \centering
     \includegraphics[width=\textwidth, trim={0.25cm 0.32cm 0.22cm 0.25cm},clip]{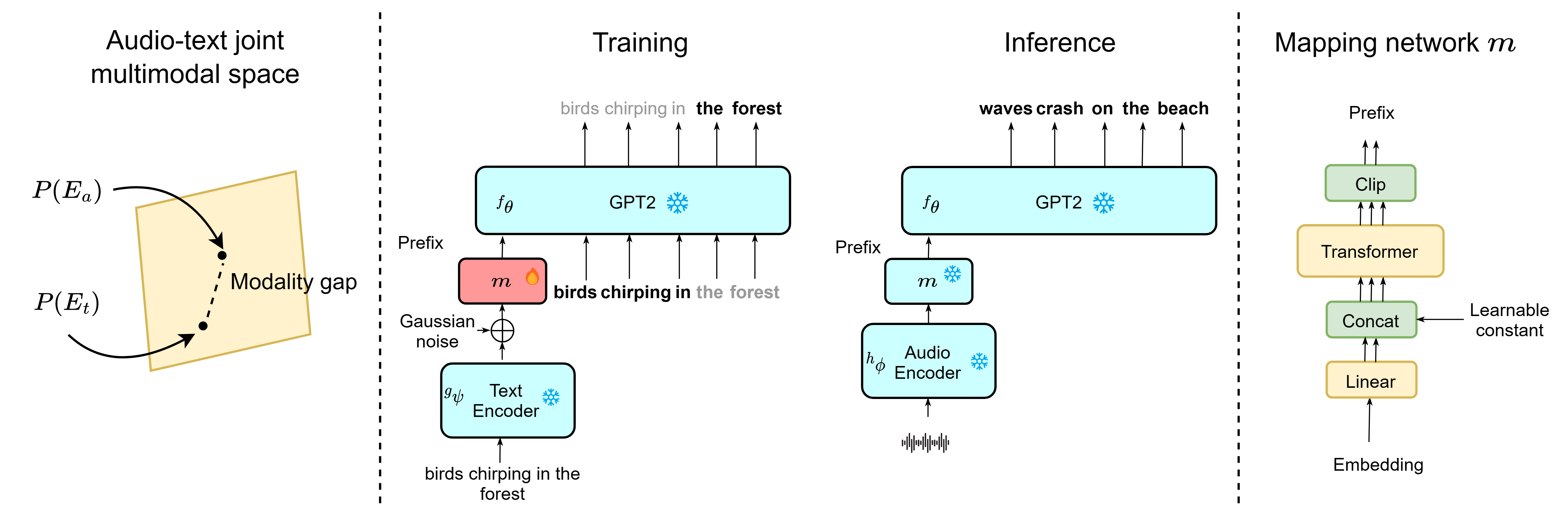}
     \caption{\small The first panel depicts the modality gap between CLAP pretrained audio and pretrained text embeddings in the joint audio-text space. The second panel shows the proposed method of text-only training for Automated Audio Captioning. At inference, the text encoder is swapped with the audio encoder and a caption is produced for the input audio. Only mapping network $m$ is trainable, while modules with \SnowflakeChevron\ are frozen. The Prefix is the output of $m$. Singular arrows depict embedding vectors while multiple arrows indicate a sequence of vectors. 
     \vspace{-0.1in}
     }
     \label{fig:clap_diagram}
\end{figure*}

Automated Audio Captioning (AAC) task involves describing the content of an audio stream in natural language. This involves describing the audio in terms of audio events, acoustic scenes, temporal relationships between events, actions, interactions between objects, and the environment. 
The typical AAC model uses an encoder-decoder architecture \cite{sutskever2014sequence} that can be semantically divided into two components: an audio understanding component and a language generation component. The audio understanding component is an audio encoder that extracts audio features. Examples of audio encoders are the pretrained sound event models like PANN \cite{pann}, AST \cite{gong2021ssast}, HTSAT \cite{chen2022hts}. The language generation component is a decoder like BART \cite{lewis-etal-2020-bart} which generates a natural language description conditioned on the audio features. 
To improve the language generation component, recent approaches use Large Language Models (LLM) with encyclopedic knowledge like GPT2 as the choice of decoder \cite{koizumi2020audio, kim2023prefix, Pengi}. However, the generated captions still suffer from short or repetitive text descriptions, of limited vocabulary and diversity \cite{martin2021diversity, kothinti2023investigations}.

Large audio-text corpora can improve AAC systems. However, collecting annotated training data for audio captioning is a challenging task: it requires human annotators to listen to each audio recording and then describe what was heard. This process is time-consuming, expensive, and introduces biases \cite{macs,kothinti2023investigations}. Recent works try to solve this problem from either a modeling or data-driven perspective. From the modeling side, researchers leverage Sound Event Detection (SED) datasets \cite{eren:2021:dcase:tech-report, Ye2021ImprovingTP,heller2023synergy}, contrastive losses \cite{liu2021cl4ac,elizalde2020never} and use training procedures that increase the diversity of captions \cite{9746894}. The data-driven approaches include scaling the audio captioning dataset by sourcing data from the web or generating data from LLM. The challenge of sourcing data from the web is the sparsity of audio-text pairs that are aligned and also human-readable. On the other hand, with LLMs, one can generate audio captions with metadata and keywords \cite{mei2023wavcaps}. However, the audio file corresponding to the generated caption is still required to train the AAC system. Therefore, we ask the question: \textit{``Can we train AAC systems using only text?"}. This way, the requirement for audio segments paired with text descriptions can be lifted during AAC training.

In this paper, we propose a method to train an AAC system using only text. Our method is based on the key insight that multimodal contrastive learning models like CLAP \cite{clap,clap2} force the audio and text embeddings in a common space.
First, we train a text decoder to generate captions conditioned on the pretrained CLAP text encoder. Then, during inference, we replace the text encoder with the pretrained CLAP audio encoder. To bridge any modality gap~\cite{liang2022mind}, we explore different lightweight approaches during training. We evaluate our method on two AAC datasets and show that our approach achieves competitive results with the state-of-the-art models trained on paired audio-text. We perform multiple ablation studies including LLM-generated text and stylized captions to verify the effectiveness of text-only training. 
\vspace{-0.1in}
\section{Approach}\label{sec:method}
\vspace{-0.1in}
CLAP \cite{clap} jointly trains an audio and text encoder using contrastive learning. After training, the audio and text encoder can be used in zero-shot setup for classification, retrieval \cite{clap, wu2022large, deshmukh2022audio} and supervised downstream tasks \cite{mei2023wavcaps, huang2023makeanaudio, liu2023audioldm, Pengi}. Our approach leverages this multimodal space to enable text-to-audio transfer. An AAC model learns $P(O|E_{a})$ where $E_{a}$ is the audio embedding and $O$ is the output caption. In our setup, $E_{a}$ is the pretrained CLAP audio embedding. As CLAP learns a joint multimodal space between audio and text, $P(E_a) = P(E_t)$ where $E_t$ is the pretrained CLAP text embedding. Therefore, for AAC, we can instead learn $P(O|E_{t}) = P(O|E_{a})$. This has two implications. First, we can use a text encoder during training and then swap it with an audio encoder during inference. Second, this enables training on text-only data without requiring aligned audio-text pairs or corresponding audio of generated caption.


\vspace{-0.1in}
\subsection{Modality Gap}
\vspace{-0.1in}
In practice, $P(E_{a}) \neq P(E_{t})$ but instead $P(E_{a}) \approx P(E_{t})$. This prevents swapping audio and text encoders as the distribution of audio and text embeddings do not exactly coincide. This effect is known as Modality Gap \cite{liang2022mind} and is shown in Fig \ref{fig:clap_diagram}. It was initially observed in image-text models like CLIP 
but applies to all multimodal contrastive learning models including CLAP. To bridge the gap in visual models, recent works have explored adding offsets and learning adapters \cite{nukrai2022text, gu2023i}. 


In this work, we explore two methods to bridge the modality gap. First, we add zero-mean Gaussian noise to CLAP's text embeddings during training. The noise helps in spreading out the text embeddings and intersecting them with the audio embeddings. The Standard Deviation $\epsilon$ of the Gaussian noise determines the robustness of the model- to the variations in embeddings and the shift caused by the modality switch. Second, we train a lightweight linear adapter to transform text embeddings into audio embeddings. We study the choice of adaptation and performance in Section \ref{sec:results}. 

\vspace{-0.05in}
\begin{table*}
\center
\footnotesize
\captionsetup{font=footnotesize}
\begin{tabular}{=l+c+c+c+c+c+c+c+c+c+c}
\hline
\makecell{Model} & \makecell{Eval. dataset} & $\text{BLUE}_1$ & $\text{BLUE}_2$ & $\text{BLUE}_3$ & $\text{BLUE}_4$ & METEOR & $\text{ROUGE}_L$ & CIDEr & SPICE & SPIDEr \\ \hline 
Chen et al. & AudioCaps & 0.489 & 0.292 & 0.178 & 0.106 & 0.152 & 0.346 & 0.265 & 0.093 & 0.179 \\
Gontier et al. & AudioCaps & 0.635 & 0.461 & 0.322 & 0.219 & 0.208 & 0.450 & 0.612 & 0.153 & 0.383 \\
Mei et al. & AudioCaps & 0.682 & 0.507 & 0.369 & 0.266 & 0.238 & 0.488 & 0.701 & 0.166 & 0.434 \\
Kim et al. & AudioCaps & 0.708 & 0.547 & 0.402 & 0.283 & 0.238 & 0.499 & 0.710 & 0.167 & 0.438 \\
Text-only (proposed)& AudioCaps & 0.645 & 0.481 & 0.338 & 0.227 & 0.220 & 0.458 & 0.697 & \textbf{0.178} & 0.437 \\ 
\rowstyle{\color{gray}} Audio-text (proposed) & AudioCaps & 0.647 & 0.480 & 0.337 & 0.223 & 0.223 & 0.462 & 0.729 & 0.181 & 0.455 \\ \hline
Chen et al. & Clotho & 0.516 & 0.325 & 0.215 & 0.141 & 0.153 & 0.350 & 0.314 & 0.102 & 0.208 \\
Gontier et al. & Clotho & 0.461 & 0.282 & 0.182 & 0.117 & 0.136 & 0.318 & 0.251 & 0.083 & 0.167 \\
Mei et al. & Clotho & 0.516 & 0.318 & 0.204 & 0.127 & 0.157 & 0.351 & 0.313 & 0.105 & 0.209 \\
Kim et al. & Clotho & 0.539 & 0.346 & 0.227 & 0.142 & 0.159 & 0.366 & 0.319 & 0.111 & 0.215 \\ 
Text-only (proposed)& Clotho & 0.524 & 0.339 & 0.222 & 0.136 & \textbf{0.173} & \textbf{0.371} & \textbf{0.379} & \textbf{0.132} & \textbf{0.256}\\ 
\rowstyle{\color{gray}} Audio-text (proposed)& Clotho & 0.574 & 0.375 & 0.250 & 0.155 & 0.173 & 0.381 & 0.398 & 0.123 & 0.261 \\ \hline
\end{tabular} 
\caption{\label{table: audio captioning results}
\small The proposed text-only method performs comparably with the best Audio Captioning models in the literature, which were trained on audio-text pairs. All models use AudioCaps and Clotho datasets in training. Higher is better for all metrics. The last two gray rows indicate model performance when audio-text pairs are used in training.} \vspace{-0.1in}
\end{table*}

\vspace{-0.1in}
\subsection{Training}
\vspace{-0.1in}
The text-only training approach and AAC model architecture are shown in Fig~\ref{fig:clap_diagram}. Let the training data in text-to-text format be referred to as \{$t^i$,$c^i$\} where $t^i$ and $c^i$ are the $i^{th}$ input and $i^{th}$ output text caption respectively. Text encoder $g_\psi$ encodes the input text $t^i$ into an embedding. We inject random Gaussian noise $\mu \sim \mathcal{N}(0, \epsilon)$ with a standard deviation of $\epsilon$ to the text embedding.
\begin{equation}
    v = g_\psi(t^i) + \mu \label{equation: prefix} 
\end{equation}
To create a prefix $p^i$, the embedding $v$ is projected to a sequence of $k$ embeddings. The prefix $p^i$ is used to prompt the pretrained frozen language model $f_\theta$. 
\begin{equation}
    p^i = p^i_1,...,p^i_{k} = m(v) \label{equation: prefix} 
\end{equation}

The language model $f_\theta$ is fed with the prefix-caption concatenation of all $\{z_i\}_{i=1}^N$,  where $z_i$ is:
\begin{equation}
    z^i = p^i_1,...,p^i_{k}, c^i_1,...,c^i_l \label{equation: prefix} 
\end{equation}
The model is trained as a standard captioning system, where it learns to predict a caption (text tokens) $c^i$ conditioned on the prefix in an autoregressive fashion. We used Cross-Entropy as the loss function:
\begin{equation}
\mathcal{L} = - \sum_{i=1}^N \sum_{j=1}^{l} \log p_{\gamma} (c^i_j| p^i_1,...,p^i_{k}, c^i_1,...,c^i_{j-1}) 
\end{equation}
where $\gamma$ denotes the model's trainable parameters, contained only within the mapping network $m$ (Figure~\ref{fig:clap_diagram}). We use a model based on prefix-tuning architecture \cite{kim2023prefix, Pengi}. However, in principle, the text-only training can be applied to any encoder-decoder AAC model. The text encoder and GPT2 are frozen.

\vspace{-0.1in}
\subsection{Inference}
\vspace{-0.1in}
At inference time, we swap the text encoder $g_{\psi}$ with the audio encoder $h_{\phi}$. We do not inject Gaussian noise or perform any modality adaptation. The audio encoder $g_\phi$ and the mapping network $m$ project the test audio $x^i$ into a sequence of $k$ embeddings.
\begin{equation}
    p^i = p^i_1,...,p^i_{k} = m(h_\phi(x^i)) \label{equation: prefix_inference} 
\end{equation}
The causal language model $f_\theta$ generates the next token sequentially conditioned on the prefix $p^i$. The language model assigns probabilities to all vocabulary tokens at each prediction, which are used to determine the next token depending on the choice of decoding. We use beam search with size 5.

\vspace{-0.1in}
\section{Experiments}\label{sec:experiments}
\vspace{-0.1in}
\textbf{Datasets.} We use text data from Clotho \cite{clotho} and AudioCaps \cite{audiocaps} for the text-only audio captioning training. In all, we use 65,002 captions for training, 24,420 captions from Clotho train set and 40,582 from AudioCaps train set. For benchmarking, we use the audios files from the test set of Clotho and test set of AudioCaps. In Section \ref{subsec:llm} we use WavCaps dataset \cite{mei2023wavcaps} to simulate text data generated by LLMs.

\noindent\textbf{Encoders.} We use the audio encoder and text encoder from CLAP~\cite{clap2}: the audio encoder is audio transformer HTSAT\cite{chen2022hts} and the text encoder is GPT2. Audio is sampled at 44.1 kHz and converted to 64-bin log Mel spectrograms, with a 1024 secs window in 50-8000 Hz, and 320 secs hop size.

\noindent\textbf{Mapper and decoder.} The mapping network $m$ shown in Figure \ref{fig:clap_diagram} consists of a linear layer, an 8-layer transformer, and a learnable constant. The prefix length is set to be 40. For the decoder, we use GPT2, specifically GPT2-base (124M parameters). The decoder is frozen through all the experiments.

\noindent\textbf{Training.} We use Adam Optimiser for 30 epochs with a linear schedule with 2000 warmup steps and a base learning rate of 1e-4. The batch size is 128 and trained on 4 V100 GPUs.
\vspace{-0.1in}
\section{Results and Discussion}\label{sec:results}
\vspace{-0.1in}
SPIDEr is used as the primary evaluation metric in line with the IEEE DCASE 2022 competition. The experiments in this section are designed to understand the utility of the text-only method (section \ref{subsec:main results}), and adapter choices (sections \ref{subsec:gaussian noise} \ref{subsec:learnable adapter}).


\begin{table*}
\center
\footnotesize
\captionsetup{font=footnotesize}
\begin{tabular}{=l+c+c+c+c+c+c+c+c+c+c}
\hline
\makecell{Model} & \makecell{Eval. dataset} & $\text{BLUE}_1$ & $\text{BLUE}_2$ & $\text{BLUE}_3$ & $\text{BLUE}_4$ & METEOR & $\text{ROUGE}_L$ & CIDEr & SPICE & SPIDEr \\ \hline
Text-only & AudioCaps & 0.645 & 0.481 & 0.338 & 0.227 & 0.220 & 0.458 & 0.696 & 0.178 & 0.437 \\
Text-only\textsuperscript{\textdagger} & AudioCaps & \textbf{0.653} & \textbf{0.484} & \textbf{0.342} & \textbf{0.232} & \textbf{0.226} & \textbf{0.459} & \textbf{0.697} & \textbf{0.179} & \textbf{0.438} \\ \hline
Text-only & Clotho & 0.524 & 0.339 & 0.222 & 0.136 & 0.173 & 0.371 & 0.379 & 0.132 & 0.256\\ 
Text-only\textsuperscript{\textdagger} & Clotho & \textbf{0.530} & \textbf{0.342} & \textbf{0.224} & \textbf{0.143} & 0.164 & 0.367 & 0.377 & 0.117 & 0.247\\ \hline
\end{tabular} 
\caption{\label{table: wavcaps}
\small Text-only uses AudioCaps and Clotho datasets in training. Symbol \textsuperscript{\textdagger} indicates LLM generated text \cite{mei2023wavcaps} is added to training data.} 
\vspace{-0.1in}
\end{table*}

\begin{table*}
\center
\footnotesize
\captionsetup{font=footnotesize}
\begin{tabular}{=l+c+c+c+c+c+c+c+c+c+c+c}
\hline
\makecell{Model} & Adapter & \makecell{Eval. dataset} & $\text{BLUE}_1$ & $\text{BLUE}_2$ & $\text{BLUE}_3$ & $\text{BLUE}_4$ & METEOR & $\text{ROUGE}_L$ & CIDEr & SPICE & SPIDEr \\ \hline
Text-only & Gaussian & AudioCaps & \textbf{0.645} & \textbf{0.481} & \textbf{0.338} & \textbf{0.227} & \textbf{0.220} & \textbf{0.458} & \textbf{0.696} & \textbf{0.178} & \textbf{0.437} \\
Text-only & $\text{Linear}_1$ & AudioCaps & 0.609 & 0.423 & 0.286 & 0.181 & 0.204 & 0.429 & 0.602 & 0.174 & 0.388 \\ \hline
Text-only & Gaussian & Clotho & 0.524 & 0.339 & 0.222 & 0.136 & \textbf{0.173} & 0.371 & 0.379 & 0.132 & 0.256\\ 
Text-only & $\text{Linear}_1$ & Clotho & \textbf{0.568} & \textbf{0.375} & \textbf{0.251} & \textbf{0.158} & 0.172 & \textbf{0.378} & \textbf{0.394} & 0.127 & \textbf{0.261}\\  \hline
\end{tabular} 
\caption{\label{table: adapter results}
\small All models use AudioCaps and Clotho datasets in training. Symbol \textsuperscript{\textdagger} indicates that LLM-generated text \cite{mei2023wavcaps} is added in training. 
} 
\vspace{-0.2in}
\end{table*}
\vspace{-0.1in}
\subsection{Text-only training}\label{subsec:main results}
\vspace{-0.1in}
Table \ref{table: audio captioning results} shows results of various models trained on both AudioCaps and Clotho. Models in rows 1-4 use both audio and text in training. The proposed text-only model (row 5) uses only text data and random Gaussian noise with a std of 0.015. It achieves comparable performance with the best audio captioning models in the literature and obtains a SPIDEr score of 0.256 on Clotho and 0.455 on AudioCaps, higher than 0.215 and 0.437 reported by Kim et. al\cite{kim2023prefix}.

Text-only training is a valid alternative to training and/or initializing audio captioning systems. We also train our model architecture made for text-only training with audio-text pairs. The architecture is similar to Fig \ref{fig:clap_diagram}, where during training we use audio files with an audio encoder instead of text with a text encoder and Gaussian noise. This is the last and grayed row in Table \ref{table: audio captioning results}. The difference in SPIDEr score between the audio-text and the text-only training is small: +0.02 on AudioCaps and +0.01 on Clotho. This indicates that our text-only training can achieve comparable results without audio data. The main benefit of text-only training is training on unpaired openly available text. We explore this in Section \ref{subsec:llm}), whereby using LLM-generated text, we show that text-only training can improve over the audio-text training.
\vspace{-0.1in}
\subsection{Gaussian Noise Injection}\label{subsec:gaussian noise}
\begin{figure}[ht]
   \centering
     \includegraphics[width=0.42\textwidth]{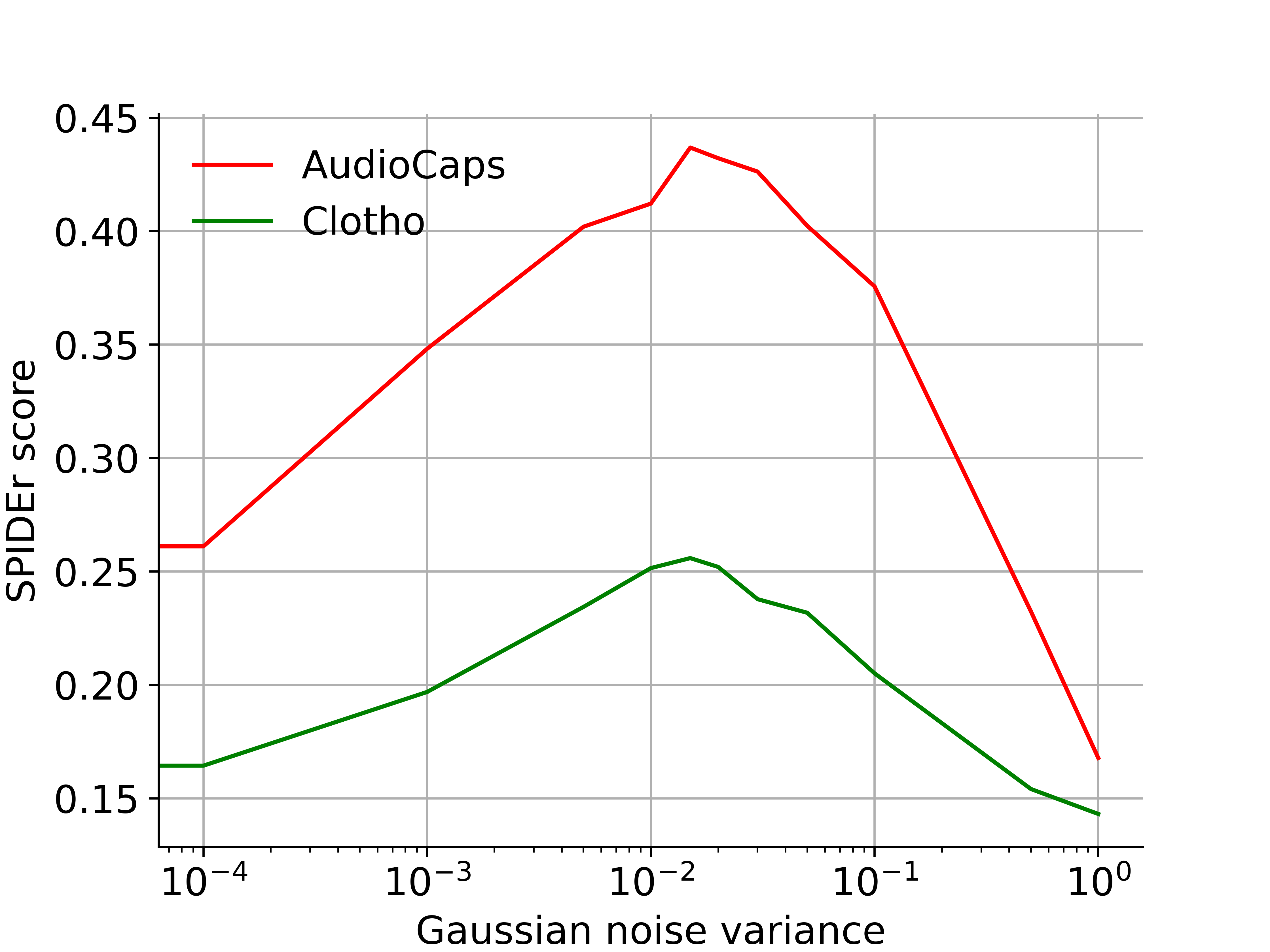}
     \caption{Effect of random Gaussian noise variance. 
     \vspace{-0.2in}}
     \label{fig:var_spider}
\end{figure}

\vspace{-0.1in}
The choice and magnitude of noise used has a significant effect on text-only training. To determine the value of variance, we take the infinity norm between the audio and text embeddings of randomly chosen 30 examples. The variance value obtained is $\sim$ 0.015. We also verify this empirically. We change the value of Gaussian noise variance from 0.0 to 1.0 and train a text-only model for each value. The SPIDEr score on AudioCaps and Clotho with respect to variance is shown in Figure \ref{fig:var_spider}. The experimental result of $\sim$ 0.015 confirms that with only 30 examples we can approximate the noise variance required for text-only training. 
%
\vspace{-0.1in}
\subsection{Learnable Adapter}\label{subsec:learnable adapter}
\vspace{-0.1in}
Random Gaussian noise is used as the default adapter in our experiments. However, with access to audio-text pairs, an adapter can be trained to bridge the modality gap. We consider learning a linear adapter to bridge the modality gap. If $f(a)$ is a pretrained audio encoder and $g(t)$ is a pretrained text encoder from CLAP. The linear adapter $h$ is applied on top of pretrained text encoder $g(t)$. The loss function is MSE between $f(t)$ and $h(g(t))$. We use AudioCaps and Clotho train-set for training the linear adapter $h$. 

Once $h$ is trained, we perform the same text-only training shown in Figure \ref{fig:clap_diagram} with an additional linear adapter applied before the Gaussian Noise. The performance with linear adapter is shown in Table \ref{table: adapter results}. There is performance improvement achieved on Clotho dataset but the performance drops on the AudioCaps dataset. A potential solution is to train an adapter on 4.3M audio-text pairs used in CLAP training. However, this deviates from our work's theme of lightweight adaptation and hence left for future work.

\vspace{-0.1in}
\section{Enriching and Stylizing captions}\label{sec:results2}
\vspace{-0.1in}


Results above show that efficient text-to-audio
transfer is attainable. And if text-only data can sufficiently train an AAC system, this has an important practical significance: additional sources of text can be used for training, including existing datasets and web data, and ``unlimited'' text from LLMs. Moreover, domain adaption can be less laborious, allowing for stylizing captions with no or limited human curation.

\vspace{-0.1in}
\subsection{Training on LLM generated text}\label{subsec:llm}
\vspace{-0.1in}
In this section, we invoke WavCaps~\cite{mei2023wavcaps} to simulate LLM-generated text. WavCaps is a publicly available dataset with text captions generated by OpenAI's ChatGPT, after utilizing human-curated metadata. We augment the training of the text-only model with the WavCaps text data, on top of AudioCaps' and Clotho's training sets.
Results, marked with \textsuperscript{\textdagger} in Table~\ref{table: wavcaps}, show performance improvements on N-gram and text matching metrics ($\text{BLUE}_i$) for both datasets.
This finding illustrates the efficient utilization of augmented text-only training data, which could also help improve vocabulary diversity.

\vspace{-0.1in}
\subsection{Stylized Audio Captioning}\label{subsec:style}
\vspace{-0.1in}
An additional consideration for text or caption descriptions is that various sources of text may also contain diverse styles. In this section we demonstrate the ability of the proposed text-only training framework to produce accurate and stylistically correct audio captions. We use Clotho and convert the original human-curated captions to a different style, ``humorous", by invoking OpenAI's GPT-4, and prompting the model to stay close to the acoustic descriptions of the original captions. For example, the original caption ``Sand is being shoveled and dumped on the ground", was transformed to ``Sand relocation program: from shovel to ground, it's a gritty story". Ideally, one can adapt the model to any snippet of text from the web. In Table~\ref{table: style results}, we show that training the text-only AAC on stylized captions is possible, and useful for model adaptation to different domains. The captions will be released\footnote{\href{https://github.com/microsoft/NoAudioCaptioning}{https://github.com/microsoft/NoAudioCaptioning}}.
%

\begin{table}[h]
\center
\footnotesize
\captionsetup{font=footnotesize}
\begin{tabular}{=l+c+c+c+c}
\hline
\makecell{Train Dataset} & \makecell{Eval. dataset} & $\text{BLUE}_1$ & $\text{BLUE}_2$ & SPIDEr \\ \hline
Original Clotho & Humor Clotho & 0.370 & 0.162 & 0.092 \\ 
Humor Clotho & Humor Clotho & \textbf{0.410} & \textbf{0.214} & \textbf{0.102}\\ \hline
\end{tabular} 
\caption{\label{table: style results}
\small Text-only AAC model trained with the original or humorous stylized captions and evaluated on the humorous styled captions.} 
\vspace{-0.2in}
\end{table}


\vspace{-0.13in}
\section{Conclusion}\label{sec:conclusions}
\vspace{-0.1in}
We introduce a method to train an AAC system on text-only data, without the need for paired audio samples. Our method uses the insight that contrastive models like CLAP force the audio and text embeddings in a common space, with some modality gap. To bridge the modality gap, we explore different light-weight approaches that can be used during training. We evaluate the proposed method on two AAC datasets and show that our text-only approach achieves competitive results with state-of-the-art audio captioning models trained on audio-text data, while it effectively allows for straightforward text-data augmentation and for stylized generated outputs.



 \begin{spacing}{0.75}
\bibliographystyle{IEEEtran}
\bibliography{refs.bib,IEEEsettings.bib}

\begin{thebibliography}{10}
\providecommand{\url}[1]{#1}
\def\UrlFont{\rmfamily}
\providecommand{\newblock}{\relax}
\providecommand{\bibinfo}[2]{#2}
\providecommand\BIBentrySTDinterwordspacing{\spaceskip=0pt\relax}
\providecommand\BIBentryALTinterwordstretchfactor{4}
\providecommand\BIBentryALTinterwordspacing{\spaceskip=\fontdimen2\font plus
\BIBentryALTinterwordstretchfactor\fontdimen3\font minus
  \fontdimen4\font\relax}
\providecommand\BIBforeignlanguage[2]{{%
\expandafter\ifx\csname l@#1\endcsname\relax
\typeout{** WARNING: IEEEtran.bst: No hyphenation pattern has been}%
\typeout{** loaded for the language `#1'. Using the pattern for}%
\typeout{** the default language instead.}%
\else
\language=\csname l@#1\endcsname
\fi
#2}}

\bibitem{sutskever2014sequence}
I.~Sutskever, O.~Vinyals, and Q.~V. Le, ``Sequence to sequence learning with
  neural networks,'' \emph{Advances in neural information processing systems},
  vol.~27, 2014.

\bibitem{pann}
Q.~Kong, Y.~Cao, T.~Iqbal, Y.~Wang, \emph{et~al.}, ``Panns: Large-scale
  pretrained audio neural networks for audio pattern recognition,''
  \emph{IEEE/ACM Trans. Audio, Speech and Lang. Proc.}, 2020.

\bibitem{gong2021ssast}
Y.~Gong, C.-I.~J. Lai, Y.-A. Chung, and J.~Glass, ``Ssast: Self-supervised
  audio spectrogram transformer,'' \emph{arXiv preprint arXiv:2110.09784},
  2021.

\bibitem{chen2022hts}
K.~Chen, X.~Du, B.~Zhu, Z.~Ma, T.~Berg-Kirkpatrick, and S.~Dubnov, ``Hts-at: A
  hierarchical token-semantic audio transformer for sound classification and
  detection,'' in \emph{IEEE International Conference on Acoustics, Speech and
  Signal Processing (ICASSP)}.\hskip 1em plus 0.5em minus 0.4em\relax IEEE,
  2022.

\bibitem{lewis-etal-2020-bart}
M.~Lewis, Y.~Liu, N.~Goyal, and et. al., ``{BART}: Denoising
  sequence-to-sequence pre-training for natural language generation,
  translation, and comprehension,'' in \emph{Proceedings of the 58th Annual
  Meeting of the Association for Computational Linguistics}.\hskip 1em plus
  0.5em minus 0.4em\relax ACL, July 2020.

\bibitem{koizumi2020audio}
Y.~Koizumi, Y.~Ohishi, D.~Niizumi, \emph{et~al.}, ``Audio captioning using
  pre-trained large-scale language model guided by audio-based similar caption
  retrieval,'' \emph{arXiv preprint arXiv:2012.07331}, 2020.

\bibitem{kim2023prefix}
M.~Kim, K.~Sung-Bin, and T.-H. Oh, ``Prefix tuning for automated audio
  captioning,'' in \emph{IEEE International Conference on Acoustics, Speech and
  Signal Processing (ICASSP)}, 2023.

\bibitem{Pengi}
S.~Deshmukh, B.~Elizalde, R.~Singh, and H.~Wang, ``Pengi: An audio language
  model for audio tasks,'' \emph{arXiv preprint arXiv:2305.11834}, 2023.

\bibitem{martin2021diversity}
I.~Martin~Morato and A.~Mesaros, ``Diversity and bias in audio captioning
  datasets,'' 2021.

\bibitem{kothinti2023investigations}
S.~Kothinti and D.~Emmanouilidou, ``Investigations in audio captioning:
  Addressing vocabulary imbalance and evaluating suitability of
  language-centric performance metrics,'' \emph{arXiv preprint
  arXiv:2211.06547}, 2023.

\bibitem{macs}
I.~Mart{\'\i}n-Morat{\'o} and A.~Mesaros, ``What is the ground truth?
  reliability of multi-annotator data for audio tagging,'' in \emph{2021 29th
  European Signal Processing Conference (EUSIPCO)}, 2021.

\bibitem{eren:2021:dcase:tech-report}
A.~O. Eren and M.~Sert, ``Audio captioning using sound event detection,''
  DCASE2021 Challenge, Tech. Rep., Jun. 2021.

\bibitem{Ye2021ImprovingTP}
Z.~Ye, H.~Wang, D.~Yang, and Y.~Zou, ``Improving the performance of automated
  audio captioning via integrating the acoustic and semantic information,'' in
  \emph{Workshop on Detection and Classification of Acoustic Scenes and
  Events}, 2021.

\bibitem{heller2023synergy}
L.~M. Heller, B.~Elizalde, B.~Raj, and S.~Deshmuk, ``Synergy between human and
  machine approaches to sound/scene recognition and processing: An overview of
  icassp special session,'' \emph{arXiv preprint arXiv:2302.09719}, 2023.

\bibitem{liu2021cl4ac}
X.~Liu, Q.~Huang, X.~Mei, T.~Ko, H.~L. Tang, M.~D. Plumbley, and W.~Wang,
  ``Cl4ac: A contrastive loss for audio captioning,'' \emph{Proceedings of the
  Detection and Classification of Acoustic Scenes and Events 2021 Workshop
  (DCASE 2021)}, 2021.

\bibitem{elizalde2020never}
B.~M. Elizalde, ``Never-ending learning of sounds,'' Ph.D. dissertation, CMU
  Pittsburgh, PA, 2020.

\bibitem{9746894}
X.~Mei, X.~Liu, J.~Sun, M.~D. Plumbley, and W.~Wang, ``Diverse audio captioning
  via adversarial training,'' in \emph{ICASSP 2022 - 2022 IEEE International
  Conference on Acoustics, Speech and Signal Processing (ICASSP)}, 2022, pp.
  8882--8886.

\bibitem{mei2023wavcaps}
X.~Mei, C.~Meng, H.~Liu, Q.~Kong, T.~Ko, C.~Zhao, M.~D. Plumbley, Y.~Zou, and
  W.~Wang, ``Wavcaps: A chatgpt-assisted weakly-labelled audio captioning
  dataset for audio-language multimodal research,'' \emph{arXiv preprint
  arXiv:2303.17395}, 2023.

\bibitem{clap}
B.~Elizalde, S.~Deshmukh, M.~A. Ismail, and H.~Wang, ``Clap learning audio
  concepts from natural language supervision,'' in \emph{IEEE International
  Conference on Acoustics, Speech and Signal Processing (ICASSP)}, 2023.

\bibitem{clap2}
B.~Elizalde, S.~Deshmukh, and H.~Wang, ``Natural language supervision for
  general-purpose audio representations,'' \textit {submitted to IEEE
  International Conference on Acoustics, Speech and Signal Processing
  (ICASSP)}, 2024.

\bibitem{liang2022mind}
W.~Liang, Y.~Zhang, Y.~Kwon, S.~Yeung, and J.~Zou, ``Mind the gap:
  Understanding the modality gap in multi-modal contrastive representation
  learning,'' in \emph{Advances in Neural Information Processing Systems},
  A.~H. Oh, A.~Agarwal, D.~Belgrave, and K.~Cho, Eds., 2022.

\bibitem{wu2022large}
Y.~Wu, K.~Chen, T.~Zhang, Y.~Hui, T.~Berg-Kirkpatrick, and S.~Dubnov,
  ``Large-scale contrastive language-audio pretraining with feature fusion and
  keyword-to-caption augmentation,'' in \emph{IEEE International Conference on
  Acoustics, Speech and Signal Processing (ICASSP)}, 2023.

\bibitem{deshmukh2022audio}
S.~Deshmukh, B.~Elizalde, and H.~Wang, ``{Audio Retrieval with WavText5K and
  CLAP Training},'' in \emph{Proc. INTERSPEECH 2023}, 2023.

\bibitem{huang2023makeanaudio}
R.~Huang, J.~Huang, D.~Yang, \emph{et~al.}, ``Make-an-audio: Text-to-audio
  generation with prompt-enhanced diffusion models,'' \emph{arXiv preprint
  arXiv:2301.12661}, 2023.

\bibitem{liu2023audioldm}
H.~Liu, Z.~Chen, Y.~Yuan, X.~Mei, \emph{et~al.}, ``Audioldm: Text-to-audio
  generation with latent diffusion models,'' \emph{arXiv preprint
  arXiv:2301.12503}, 2023.

\bibitem{nukrai2022text}
D.~Nukrai, R.~Mokady, and A.~Globerson, ``Text-only training for image
  captioning using noise-injected clip,'' in \emph{Findings of the Association
  for Computational Linguistics: EMNLP 2022}, 2022.

\bibitem{gu2023i}
S.~Gu, C.~Clark, and A.~Kembhavi, ``I can't believe there's no images! learning
  visual tasks using only language data,'' \emph{arXiv preprint
  arXiv:2211.09778}, 2023.

\bibitem{clotho}
K.~Drossos, S.~Lipping, and T.~Virtanen, ``Clotho: an audio captioning
  dataset,'' in \emph{IEEE International Conference on Acoustics, Speech and
  Signal Processing (ICASSP)}, 2020.

\bibitem{audiocaps}
C.~D. Kim, B.~Kim, H.~Lee, and G.~Kim, ``{AudioCaps: Generating Captions for
  Audios in The Wild},'' in \emph{NAACL-HLT}, 2019.

\end{thebibliography}
 \end{spacing}
\end{document}